\documentclass[aps,prd,preprint,superscriptaddress,tightenlines,nofootinbib]{revtex4}

\usepackage{graphicx}
\usepackage{dcolumn}
\usepackage{bm}

\begin{document}

\preprint{CLEO 06-10}         

\title{Branching Fraction for the Doubly-Cabibbo-Suppressed Decay 
$D^{+} \rightarrow K^{+} \pi^{0}$\hspace*{2mm}}
\thanks{Submitted to the 33$^{\rm rd}$ International Conference on High Energy
Physics, July 26 - August 2, 2006, Moscow}
 
\author{S.~A.~Dytman}
\author{W.~Love}
\author{V.~Savinov}
\affiliation{University of Pittsburgh, Pittsburgh, Pennsylvania 15260}
\author{O.~Aquines}
\author{Z.~Li}
\author{A.~Lopez}
\author{S.~Mehrabyan}
\author{H.~Mendez}
\author{J.~Ramirez}
\affiliation{University of Puerto Rico, Mayaguez, Puerto Rico 00681}
\author{G.~S.~Huang}
\author{D.~H.~Miller}
\author{V.~Pavlunin}
\author{B.~Sanghi}
\author{I.~P.~J.~Shipsey}
\author{B.~Xin}
\affiliation{Purdue University, West Lafayette, Indiana 47907}
% ...
\collaboration{CLEO Collaboration} 
\noaffiliation

\date{July 25, 2006}

\begin{abstract} 
We present a measurement of the branching fraction for the 
doubly-Cabibbo-suppressed decay 
$D^{+} \rightarrow K^{+} \pi^{0}$, using 281 $\mathrm{pb}^{-1}$ of data 
accumulated with the CLEO-c detector on the
$\psi$(3770) resonance. We find 
${\cal B} \left( D^{+} \rightarrow K^{+}
\pi^{0}\right) = (2.25 \pm 0.36 \pm 0.15 \pm 0.07) \times
10^{-4}$, 
where the first uncertainty is statistical, the second is systematic, 
and the last error is due to the uncertainty in the reference mode.
The results presented in this document are preliminary.
\end{abstract}

\maketitle

The Cabibbo-favored hadronic decays of the $c$ quark proceed 
through $c \rightarrow s
W^{+}_{V},\ W^{+}_{V} \rightarrow u \overline d$. The
doubly-suppressed decays proceed through $c \rightarrow d W^{+}_{V}$,
$W^{+}_{V} \rightarrow u \overline s$, and are expected to be suppressed
by a factor
$|(V_{cd}V_{us})/(V_{cs}V_{ud})|^{2} \approx 2.5 \times 10^{-3}$.
We have measured the
branching fraction for the doubly-Cabibbo-suppressed decay $D^{+}
\rightarrow K^{+} \pi^{0}$ (charge conjugate mode $D^{-}
\rightarrow K^{-} \pi^{0}$ implied also, except where noted).

For this measurement we have used a sample of 281 $\mathrm{pb}^{-1} e^{+}
e^{-} \rightarrow \psi(3770)$ events, produced
with the CESR-c storage ring and detected with the CLEO-c
detector. The data sample contains about $0.8 \times 10^{6}$ $D^{+}
D^{-}$ events (our target sample), one million $D^{0} \overline D^{0}$ 
events, five million $e^{+} e^{-} \rightarrow u \overline u,\ d \overline d,$ or
$s \overline s$ continuum events, one million $e^{+} e^{-}
\rightarrow \tau^{+} \tau^{-}$ events, one million $e^{+} e^{-}
\rightarrow \gamma \psi^{\prime}$ radiative return events
(sources of background), as well as Bhabha events, $\mu$-pair
events, $\gamma \gamma$ events (useful for luminosity
determination and resolution studies).

The CLEO-c detector is a general purpose solenoidal detector which
includes a tracking system for measuring momenta and specific
ionization ($dE/dx$) of charged particles, a Ring Imaging Cherenkov detector
(RICH) to
aid in particle identification, and a CsI calorimeter for detection of
electromagnetic showers. The CLEO-c detector is described in detail
elsewhere \cite{CLEODeter001,CLEODeter002,CLEODeter003,CLEODeter004}.

The resonance $\psi$(3770) is below the threshold
for $D \overline D \pi$ production, and so the events of interest, $e^{+}
e^{-} \rightarrow \psi(3770) \rightarrow D \overline D$,
have $D$ mesons with energy equal to the beam energy, and a unique
momentum. Having picked the particles being considered to make up
a $D$ meson, following Mark III \cite{MARK3}  we define the two
variables: 

\begin{eqnarray}
\Delta E \equiv \sum_{i} E_{i} - E_{\rm beam} ~~~,
\end{eqnarray}
and
\begin{eqnarray}
M_{\mathrm {bc}}\equiv \sqrt{E^{2}_{\rm beam} - |\sum_{i} \vec{P}_{i} |^{2}} ~~~,
\end{eqnarray}
where \noindent $E_{i},\ \vec{P}_{i}$ are the energy and momentum of each particle
making up the $D$ meson. For a correct combination of particles,
$\Delta E$ will be consistent with zero, and the beam-constrained mass
$M_{\mathrm {bc}}$ will be consistent with the $D$ mass.

In addition to $D^{+} \rightarrow K^{+} \pi^{0}$, we have studied
the singly-Cabibbo-suppressed decay $D^{+} \rightarrow \pi^{+}
\pi^{0}$, as a higher rate decay possessing many of the features
of $D^{+} \rightarrow K^{+} \pi^{0}$, and the Cabibbo-favored
decay $D^{+} \rightarrow K^{-} \pi^{+} \pi^{+}$, as a high rate,
clean mode used for normalization. We distinguish between
$K^{\pm}$ and $\pi^{\pm}$ using information from the RICH, and
$dE/dx$ information from the central drift chamber. We detect
$\pi^{0}$'s via $\pi^{0} \rightarrow \gamma \gamma$, detecting the
$\gamma$ rays in the CsI calorimeter. We require that the
calorimeter clusters be above 30 MeV, have a lateral distribution consistent with
that from $\gamma$ rays, and not be matched to charged tracks. We
require that the $\gamma \gamma$ invariant mass be within 3
standard deviations of the $\pi^{0}$ mass.

We select candidate combinations that have $\Delta E$ between
$-$40 MeV and +35 MeV, for $K^{+} \pi^{0}$ and $\pi^{+} \pi^{0}$,
and between $-$20 MeV and +20 MeV for $K^{-} \pi^{+} \pi^{+}$.
These are roughly 3 standard deviations requirements, based on Monte Carlo
studies. The asymmetric cut for $K^{+} \pi^{0}$ and $\pi^{+}
\pi^{0}$ is due to a low-side tail on $\pi^{0}$ energies, and the
wider window due to poorer energy resolution. To study background,
we select combinations with $\Delta E$ between $-$100 and $-$50 MeV,
and between +45 and +100 MeV (+50 and +100 MeV for $K^{-} \pi^{+}
\pi^{+}$). On the rare occasion when an event contains more than one
$K^+ \pi^0$ combination that passes our $\Delta E$ requirement, we
choose the combination with $\Delta E$ value closest to zero. Multiple
candidates per event for $\pi^+ \pi^0$ and for $K^- \pi^+ \pi^+$ are
removed by the same procedure. Thus, we allow only one candidate per
event per decay mode per $D$ charge.

\begin{figure}[htbp]

\begin{minipage}{5cm}
  \centerline{\includegraphics[height=0.25\textheight, width=5cm]
             {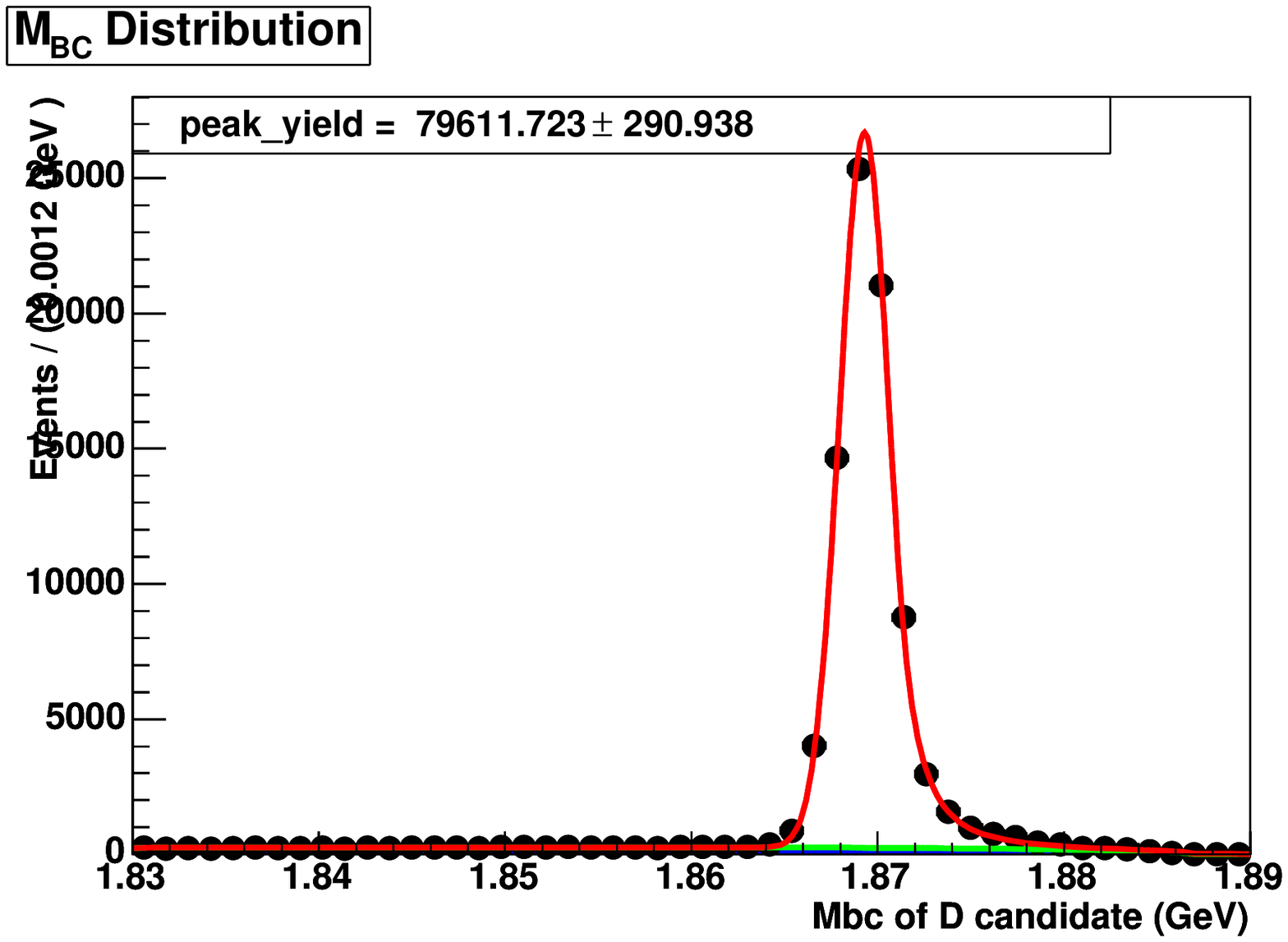}}
  \centerline{ $D^+ \rightarrow K^- \pi^+ \pi^+$}
\end{minipage}
\begin{minipage}{5cm}
  \centerline{\includegraphics[height=0.25\textheight, width=5cm]
             {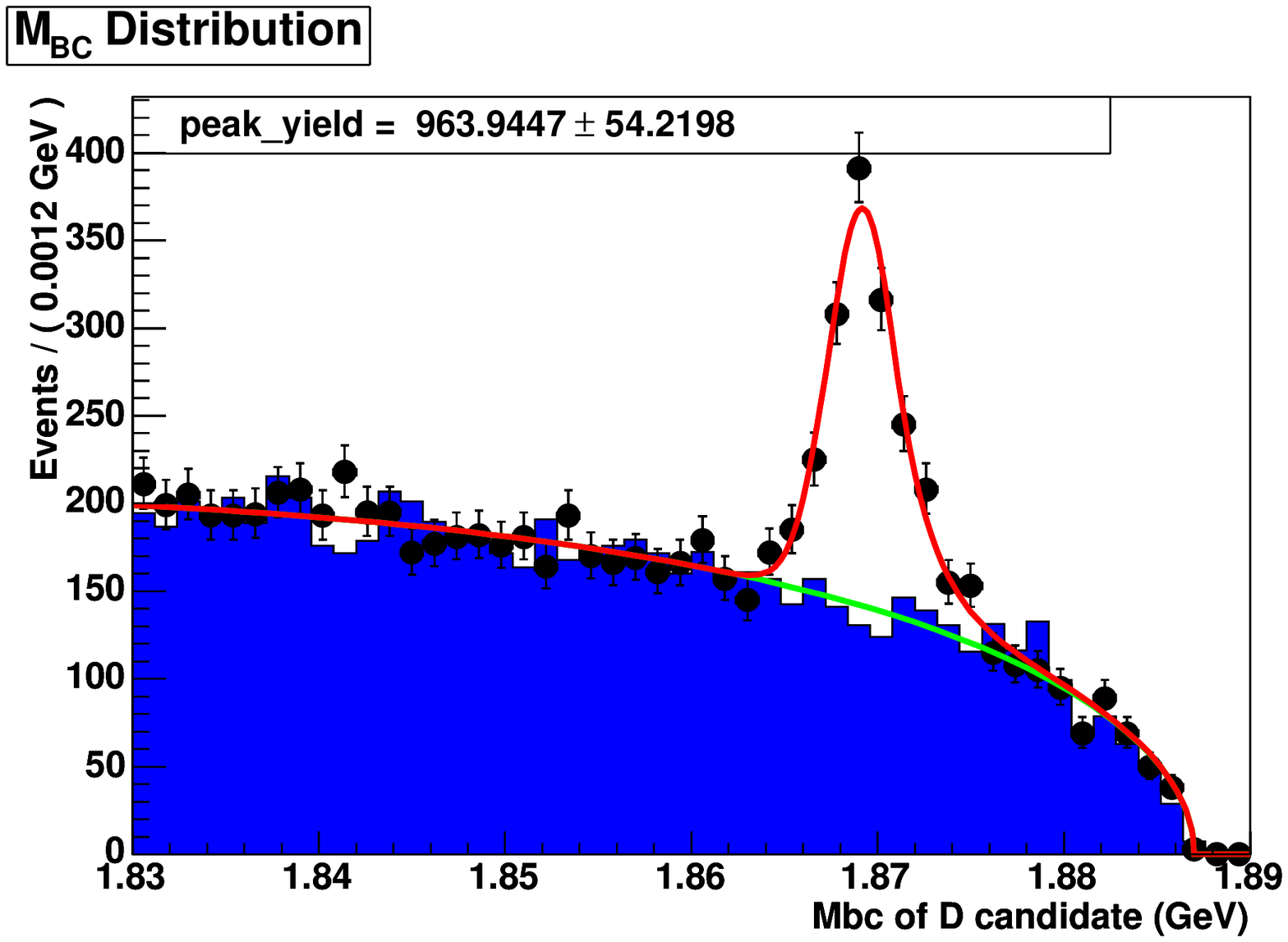}}
  \centerline{$D^+ \rightarrow \pi^+ \pi^0$ }
\end{minipage}
\begin{minipage}{5cm}
  \centerline{\includegraphics[height=0.25\textheight, width=5cm]
             {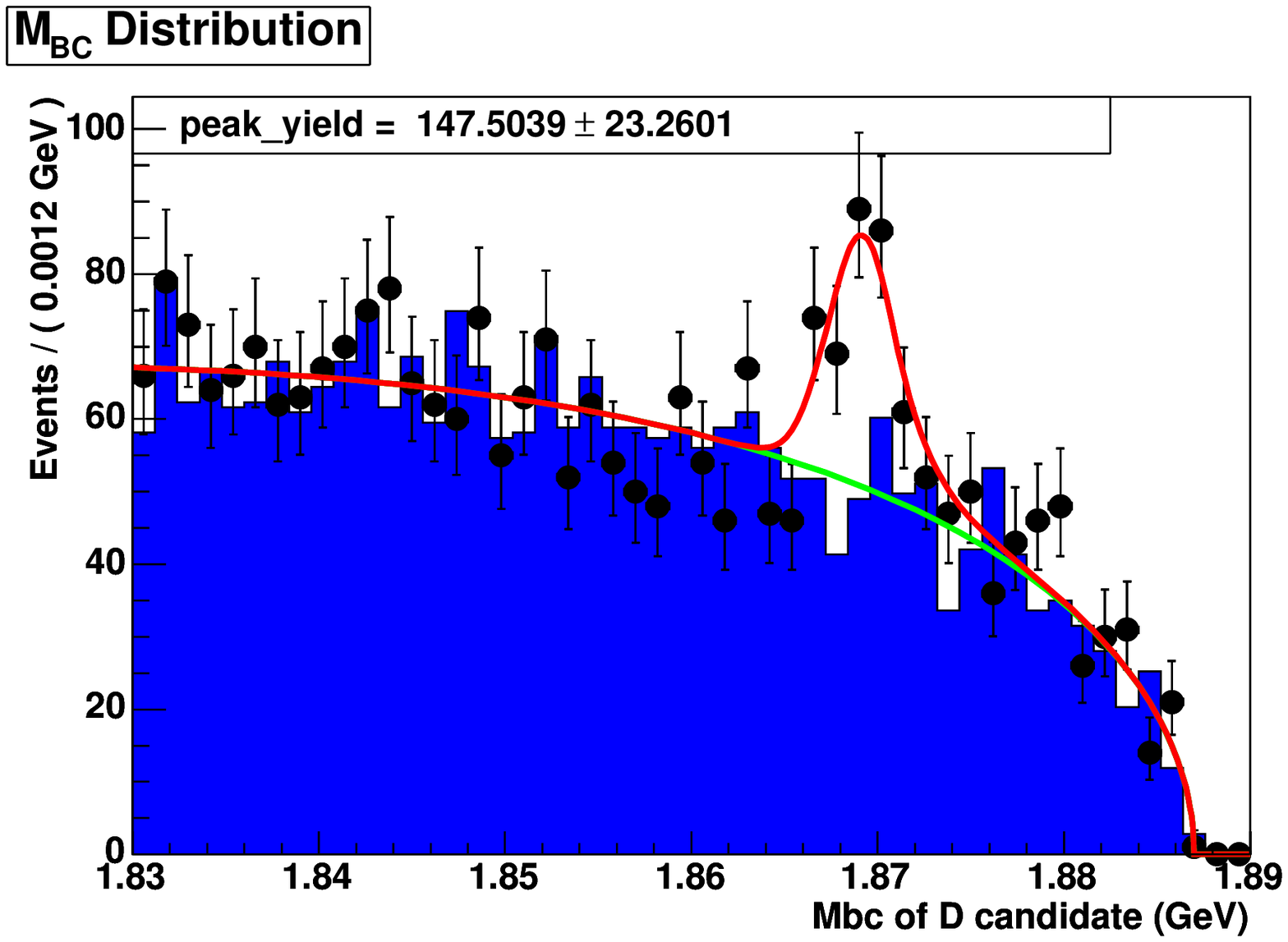}}
  \centerline{$D^+ \rightarrow K^+ \pi^0$ }
\end{minipage}
\caption{$M_{\mathrm {bc}}$ distributions of $D^+ \rightarrow K^- \pi^+ \pi^+$,  
         $D^+ \rightarrow \pi^+ \pi^0$ and 
         $D^+ \rightarrow K^+ \pi^0$. The points
         are obtained by selecting the $\Delta E$ signal region, the shaded
         histogram is from the $\Delta E$ sidebands, and the line is
         the fit described in the text. } 
\label{Fig:ST_DATA_Result}
\end{figure}


The $M_{\mathrm {bc}}$ distributions for candidate combinations are shown in
Fig. $\ref{Fig:ST_DATA_Result}$. The normalization mode $D^{+} \rightarrow K^{-} \pi^{+}
\pi^{+}$ is essentially background free. The $D^{+} \rightarrow \pi^{+}
\pi^{0}$ mode background is well described by the distribution
obtained from the $\Delta E$ sideband, as is that for the $D^{+}
\rightarrow K^{+} \pi^{0}$ mode. There is a clear peak in $D^{+}
\rightarrow K^{+} \pi^{0}$.

We perform an unbinned maximum likelihood fit to extract signal yields
from the $M_{\mathrm {bc}}$ distributions. For the background, we use an
ARGUS function \cite{ArgusFunc}, 
with shape parameter determined from the $\Delta
E$ sideband $M_{\mathrm {bc}}$ plot. For the signal, we use a Crystal Ball
line shape \cite{CBFunc}, 
which is a Gaussian with a high-side tail. As Monte
Carlo studies show that $D^{+} \rightarrow K^{+} \pi^{0}$ and
$D^{+} \rightarrow \pi^{+} \pi^{0}$ have the same signal shapes, we
have determined the line shape parameters (Gaussian peak location,
Gaussian width, point at which high-side tail begins) from the
$D^{+} \rightarrow \pi^{+} \pi^{0}$  $M_{\mathrm {bc}}$ distribution, and used
them in the fit to the $D^{+} \rightarrow K^{+} \pi^{0}$  $M_{\mathrm {bc}}$
distribution.  We have varied the shape of the high-side tail, as
part of the systematic error study.

\begin{table}[hbtp]
\centering
\caption{The MC efficiencies, fit yields and branching fractions. 
         A correction has been applied to the branching fraction
         for $\pi^0$-finding efficiency (see below).
         Only statistical uncertainties are included. }
\vspace{0.3cm}
\begin{tabular}{||c||c|c|c||}
\hline \hline
Mode                              &  $\epsilon$ (\%)   &  Yield             &  $\mathcal{B}$ (\%)        \\ \hline \hline
$D^+ \rightarrow K^- \pi^+ \pi^+$ &  54.91 $\pm$ 0.17  &  79612 $\pm$ 290   &  Input                 \\ \hline
$D^+ \rightarrow \pi^+ \pi^0$     &  50.16 $\pm$ 0.16  &    963 $\pm$ 54    &  0.1311 $\pm$ 0.0074 \\ \hline
$D^+ \rightarrow K^+ \pi^0$       &  44.53 $\pm$ 0.15  &    147 $\pm$ 23    &  0.0225 $\pm$ 0.0036 \\ 
\hline \hline
\end{tabular}
\label{Tab:ST_DATA_Result}
\end{table}

Results of the fits are shown in Table $\ref{Tab:ST_DATA_Result}$. 
Also given in Table $\ref{Tab:ST_DATA_Result}$ is
the detection efficiency for each mode, and the branching fractions
obtained for $D^{+} \rightarrow \pi^{+} \pi^{0}$ and $D^{+}
\rightarrow K^{+} \pi^{0}$. Those branching fractions are obtained
relative to $D^{+} \rightarrow K^{-} \pi^{+} \pi^{+}$, taking that
branching fraction as $(9.40 \pm 0.30)$\%, which is obtained from a 
weighted average
of the Particle Data Group (PDG) value \cite{PDGValue}
and the recent CLEO
measurement \cite{QingPaper}. The branching fraction for $D^{+}
\rightarrow \pi^{+} \pi^{0}$ is in good agreement with our
previously-published branching fraction using the same data set
($(0.125 \pm 0.006 \pm 0.007 \pm 0.004)$\%) \cite{BluskPaper}, 
is {\it not} independent of
it, and should {\it not} be used in place of it.

We have considered many sources of systematic error to the $D^{+}
\rightarrow K^{+} \pi^{0}$ branching fraction, including: signal
Monte Carlo statistics, track-finding efficiency,
$\pi^{0}$-finding efficiency, particle identification, the $\Delta
E$ requirement, final state radiation, and the uncertainty from
our fitting procedure (background shape, signal shape). The only
ones greater than 1/10 the statistical error are $\pi^{0}$-finding
efficiency, background shape, and signal shape.

The Monte Carlo simulation of the calorimeter response to photons is
imperfect, particularly in those angular regions where there is
considerable material between the interaction point and the
calorimeter. Consequently, the Monte Carlo simulation slightly
overestimates the efficiency for detecting $\pi^0$'s. Various
data-Monte Carlo comparisons suggest a correction factor of (0.95
$\pm$ 0.04), which we apply.

The background shape is determined by a fit to the $\Delta E$
sideband data. The error on the shape parameter thus determined
translates into a $\pm$4.4\% relative error in the $D^{+}
\rightarrow K^{+} \pi^{0}$ branching fraction. The signal shape is
determined by a fit to the $D^{+} \rightarrow \pi^{+} \pi^{0}$
signal.  Uncertainty comes from the determination of Gaussian
width $(\sigma)$, and point at which non-Gaussian tail sets in
$(\alpha)$. We have determined the error ellipse in the
determination of these two parameters, and noted the variation in
fitted $D^{+} \rightarrow K^{+} \pi^{0}$ yield as one travels
around this error ellipse. In that way, we obtain a systematic
error of $\pm$2.6\%, relative. Note that both the background shape
uncertainty and signal shape uncertainty are really statistical
errors, hence can be reliably determined and will decrease as
additional data are taken.

Our final result is

$${\cal B} \left( D^{+} \rightarrow K^{+}
\pi^{0}\right) = (2.25 \pm 0.36 \pm 0.15 \pm 0.07) \times
10^{-4}$$

\noindent where the first error is statistical, the second error
is systematic, and the third error is from the uncertainty in the
$D^{+} \rightarrow K^{-} \pi^{+} \pi^{+}$ branching fraction,
(9.40 $\pm$ 0.30)\%, used as the normalizing mode.

Our result is in good agreement with the only other measurement of
this branching fraction, Babar's recent ${\cal B} \left( D^{+}
\rightarrow K^{+} \pi^{0} \right) = (2.52 \pm 0.47 \pm 0.25 \pm 0.08)
\times 10^{-4}$ \cite{babar}. It can be converted to a width, using the PDG
value for the $D^{+}$ lifetime, and compared  with the width for
doubly-Cabibbo-suppressed $D^{0}$ decay $D^{0} \rightarrow K^{+}
\pi^{-}$, using the PDG value for $D^{0} \rightarrow K^{+}
\pi^{-}$ branching fraction and $D^{0}$ lifetime
\cite{PDGValue}.  In this way we obtain

\begin{eqnarray*}
\frac {\Gamma(D^{+} \rightarrow K^{+} \pi^{0})} {\Gamma(D^{0}
\rightarrow K^{+} \pi^{-})} = \frac 
{{\cal B} (D^{+} \rightarrow K^{+} \pi^{0}) \times \tau_{D^0}} 
{{\cal B} (D^{0} \rightarrow K^{+} \pi^{-}) \times \tau_{D^+}}
 =0.64 \pm 0.12
\end{eqnarray*}

\noindent 
Implications of our result are discussed in~\cite{CLEO_D_K0pi}.

We gratefully acknowledge the effort of the CESR staff 
in providing us with excellent luminosity and running conditions. 
D.~Cronin-Hennessy and A.~Ryd thank the A.P.~Sloan Foundation. 
This work was supported by the National Science Foundation,
the U.S. Department of Energy, and 
the Natural Sciences and Engineering Research Council of Canada.


\begin{thebibliography}{99}
\bibitem{CLEODeter001} G. Bonvicini $et ~ al$. (CLEO Collaboration), 
                       Phys. Rev. D \textbf{70}, 112004 2004 [hep-ex/0411050].
\bibitem{CLEODeter002} Y. Kubota $et ~ al$. (CLEO Collaboration), Nucl. Instrum. and
                       Meth. A \textbf{320}, 66 (1992).
\bibitem{CLEODeter003} D. Peterson $et ~ al$., Nucl. Instrum. and Meth.
                       A \textbf{478}, 142 (2002).
\bibitem{CLEODeter004} M. Artuso  $et ~ al$., Nucl. Instrum. and Meth.
                       A \textbf{502}, 91 (2003).
\bibitem{MARK3}        J. Adler $et ~ al$. (Mark III Collaboration),
                       Phys. Rev. Lett. \textbf{62}, 1821(1989).
\bibitem{ArgusFunc}    H. Albrecht $et ~ al$. (ARGUS Collaboration), Phys. Lett B \textbf{229},
                       304 (1989).
\bibitem{CBFunc}       T. Skwarnicki, Ph.D thesis, Institute for
                       Nuclear Physics, Krakow, Poland, 1986.
\bibitem{QingPaper}    Q. He $et ~ al$. (CLEO Collaboration),
                       Phys. Rev. Lett. \textbf{95}, 121801 (2005).
\bibitem{BluskPaper}   P.Rubin $et ~ al$. (CLEO Collaboration),
                       Phys. Rev. Lett. \textbf{96}, 081802 (2006).
\bibitem{babar}        B. Aubert $et ~ al$. (BaBar Collaboration),
                       Phys. Rev. D {\bf 74}, 011107(R) (2006).
\bibitem{PDGValue}     S. Eidelman $et ~ al$., Phys. Lett. B \textbf{592}, 1 (2004).
\bibitem{CLEO_D_K0pi}  Q. He $et ~ al$. (CLEO Collaboration), CLEO-CONF-06-11,
                       contributed to ICHEP06 [arXiv:hep-ex/0607068].

\end{thebibliography}
\end{document}